\documentclass[elsarticle,amsmath,amssymb,floatfix]{revtex4-1}

\pdfoutput=1

\usepackage{scrpage2}			
\usepackage{graphicx}			
\usepackage{amsmath}			
\usepackage{amsfonts}
\usepackage{amssymb}
\usepackage{amsthm}
\usepackage{xcolor}
\usepackage{subfigure}
\usepackage{hyperref}

\newcommand{\bqa}{\begin{eqnarray}}
\newcommand{\eqa}{\end{eqnarray}}
\newcommand{\nn}{\nonumber \\}
\newcommand{\beq}{\begin{equation}}
\newcommand{\eeq}{\end{equation}}

\def\be{\begin{eqnarray}}
\def\ee{\end{eqnarray}}

\begin{document}

\title{Exceptional point description of one-dimensional chiral topological superconductors/superfluids in BDI class}

%
%

 \author{Ipsita Mandal}
 \address{Perimeter Institute for Theoretical Physics, 31 Caroline St. N., Waterloo ON N2L 2Y5, Canada}
 \email{imandal@pitp.ca}
 \author{Sumanta Tewari}
 \address{Department of Physics and Astronomy, Clemson University, Clemson, SC 29634, USA}
 \email{stewari@g.clemson.edu}
\date{\today}

\begin{abstract}
We show that certain singularities of the Hamiltonian in the complex wave vector space can be used to identify topological quantum phase transitions for  $1D$ chiral topological superconductors/superfluids in the BDI class. These singularities fall into the category of the so-called exceptional points ($EP$'s) studied in the
context of non-Hermitian Hamiltonians describing open quantum systems. We also propose a generic formula in terms of the properties of the $EP$'s to quantify the exact number of Majorana zero modes in a particular chiral topological superconducting phase, given the values of the parameters appearing in the Hamiltonian. This formula serves as an alternative to the familiar integer ($\mathbb{Z}$) winding number invariant characterizing topological superconductor/superfluid phases in the chiral BDI class.
\end{abstract}

\maketitle

\section{Introduction}
Exceptional points ($EP$'s) are singular points in the parameter space of an operator at which two or more of its eigenvalues and eigenvectors coalesce \cite{Panch,Berry,Kato,Rotter,Heiss,Fagotti,Brand}. If the
operator is the Hamiltonian itself, $EP$'s describe collapse of two or more energy eigenvalues at certain points in the parameter space similar to a degeneracy point, but with the important difference that the energy eigenvectors are not orthogonal to each other. The degeneracy of the eigenvalues with concomitant degeneracy of the eigenvectors gives rise to a whole host of non-trivial phenomena. For this reason, $EP$'s have recently attracted enormous interest in the literature \cite{Berry,Kato,Rotter,Heiss,Fagotti,Brand,Berry2,Moise,sourin,ramon,dmitry1,dmitry2}. Although the concept of $EP$'s is known in mathematics for many years, their application in physics has been mostly limited to open quantum
systems with dissipation, appropriately described by non-Hermitian Hamiltonians \cite{Berry2,Moise,Brand}.

Topological superconductors\cite{Hasan_2010} are systems characterized by a bulk superconducting gap, and, topologically protected zero energy edge states known as MBSs (Majorana Bound States) described by second quantized operators satisfying the operator relation $\gamma^{\dagger}=\gamma$. In the context of condensed matter physics, aside from being fascinating emergent non-elementary particles (which can be identified with their own anti-particles), MBSs obey Ising type non-Abelian braiding statistics \cite{Kitaev_2001,Kitaev_2003} potentially useful in implementing a fault-tolerant topological quantum computer \cite{Kitaev_2001,Nayak_2008}. While MBSs have not yet been conclusively found in nature, they have been theoretically shown to exist in low dimensional spinless $p$-wave superconducting systems \cite{Read_Green_2000,Kitaev_2001}, as well as other systems involving various heterostructures with proximity-induced superconductivity which are topologically similar to them \cite{Fu_2008,Zhang_2008,Sau,Long-PRB,Roman,Oreg}.
In particular, the spin-orbit coupled semiconductor-superconductor heterostructure scheme has motivated tremendous experimental efforts with a number of recent works claiming to have observed experimental signatures of MBSs in zero bias tunneling experiments\cite{Mourik_2012,Deng_2012,Das_2012,Finck_2013}. More recently, experiments on ferromagnetic Fe-atom chains embedded on Pb superconductor substrate, have also seen tantalizing evidence of MBSs in spatially resolved scanning tunneling microscopy measurements \cite{Yazdani_2014}.

Recent theoretical work \cite{Schnyder_2008,Ryu_2010} has established that the quadratic Hamiltonians for gapped topological insulators and topological superconductors can be classified into ten topological symmetry classes, each of which is characterized by a topological invariant.
The symmetry classification is important as it provides an understanding of the effects of various perturbations on the stability of the protected surface modes, such as MBSs. The (strictly $1D$) semiconductor-superconductor nanowire structure \cite{Long-PRB,Roman,Oreg}, as well as the system of ferromagnetic atomic chains or nanowires deposited on Pb superconductor \cite{Yazdani_2014,Hui_2014,sumanta-chiral}, are in the topological class BDI. They are also known as chiral topological superconductors, described by an integer ($\mathbb{Z}$) winding number topological invariant \cite{TewariPRL2012}
that counts the number of protected zero energy Majorana modes at the individual edges. While the $1D$ semiconductor-superconductor nanowire structure is topologically isomorphic
to the $1D$ spinless $p$-wave superconductor or Kitaev model \cite{Kitaev_2001} (which is in the chiral BDI class in the absence of symmetry breaking perturbations \cite{TewariPRL2012,Fidkowski,tewari_PRB_2012}), the system of ferromagnetic atomic chain or nanowire embedded on Pb superconductor is isomorphic to the doubled (or time reversal symmetric) Kitaev model \cite{Dumitrescu_TR_Symm}, which is also in the chiral BDI class with a $\mathbb{Z}$ invariant. In the presence of chiral symmetry breaking terms (say, for example, stray magnetic fields and/or magnetic impurities), the symmetry classification of these systems reduces to class D, described by a $\mathbb{Z}_2$ invariant, and the number of protected MBSs at any given end reduces to zero or one.

The topological phases and quantum phase transitions within the BDI class topological superconductors are usually described in terms of the closing and re-opening of the single particle energy gap, and a winding number integer topological invariant associated with the bulk Hamiltonian with periodic boundary conditions that counts the number of protected zero energy end states localized at any given edge \cite{Schnyder_2008,Ryu_2010,TewariPRL2012}. In this paper, we describe the topological quantum phase transitions in the BDI class, and propose a generic formula to count the exact number of Majorana zero modes in $1D$ chiral topological superconductors/superfluids based on the notion of $EP$'s. This quantity can exactly point out which topological phase we are considering, depending on the values of the parameters. In a recent work \cite{sourin}, the notion of the $EP$'s was discussed in the context of Majorana zero modes for the $1D $
Kitaev model \cite{Kitaev_2001}.

Corresponding to a physical $1D$ $2N \times 2N $ Hamiltonian for a chiral topological superconductor, one can construct a non-Hermitian matrix by complexifying the momentum $k$. In the chiral basis, the Hamiltonian has two $N \times N$ off-diagonal blocks, which is a property of a chiral topological superconducting system. The values of the complex $k$, where the determinant of any one of the two off-diagonal elements vanish, are points in the complex $k$ space where two (or more) eigenvalues of the complexified Hamiltonian vanish, and will be an example of the $EP$'s. These singularities are those special points where two or more repelling levels are connected by a square root branch point in the complex $k$-plane \cite{Heiss}. More details can be found in Appendix~\ref{app:append}. We will see that one or more $EP$'s collapse at the points of a topological quantum phase transitions in the parameter space. In other words, the determinants of both the off-diagonal blocks reduce to zero at the phase transition points. In the complex $k$-plane, if one expands the off-diagonal Hamiltonian around a solution $k=k_{EP}$ for an $EP$, the zeroth order piece is non-diagonalizable, as it consists of the determinant of one of the off-diagonal elements going to zero. The leading order correction is off-diagonal with determinant of both the blocks non-zero, and the entire matrix with these two non-zero off-diagonal blocks being diagonalizable. At a topological phase transition point, the the determinants of both the blocks of the zeroth order matrix vanishes, and we are left with a completely diagonalizable Hamiltonian on expanding about that value of $k_{EP}$. Elucidating the nature of a topological quantum phase transition in terms of $EP$'s, we propose a generic formula in terms of the properties of the $EP$'s to quantify the exact number of Majorana zero modes in a given chiral topological superconducting phase, given the values of the parameters appearing in the Hamiltonian. This formula serves as an alternative to the familiar integer ($\mathbb{Z}$) winding number invariant \cite{Schnyder_2008,Ryu_2010,TewariPRL2012} characterizing topological superconductor/superfluid phases in the chiral BDI class.

The paper is organized as follows: In Sec.~\ref{model1}, we consider the Kitaev model \cite{Kitaev_2001} of $1D$ spinless $p$-wave superconductor. Sec.~\ref{model2} is devoted to the study of the quantum Ising chain with longer-ranged interactions \cite{sudip,diptiman1,diptiman2}, which, by a Jordan-Wigner transformation, maps on to the $1D$ Kitaev model with longer range hopping and superconducting pair potential.
We study a third model of the Majorana fermions in chiral topological ferromagnetic nanowires, with proximity-induced superconductivity \cite{Hui_2014,sumanta-chiral}, in Sec.~\ref{model3}. All the three systems, for appropriate values of the parameters, can support chiral Majorana bound states at any given end. In Sec.~\ref{gen-formula}, we propose a generic formula to count the total number of zero modes in a particular chiral topological phase, given the values of the parameters of the Hamiltonian. We also provide the mathematical proof of why this quantity is related to the different topological phases. Lastly, we finish with a summary and outlook in Sec.~\ref{conclusion}. 
In Appendix~\ref{app:append}, we review the definition of $EP$'s.


\section{Model $1$: Kitaev Chain}
\label{model1}

Kitaev's model of $1D$ p-wave superconducting quantum wire can support Majorana zero modes at the ends \cite{Kitaev_2001}, depending on the value of the chemical potential $\mu$.
In the Bogoliubov-de Gennes (BdG) basis, the Hamiltonian is given by:
\begin{eqnarray}
H_1 (k)  &=&  \left(\cos(k) - \mu \right) \,
\sigma_y - \Delta \, \sin(k) \, \sigma_x\,,
\label{h1}
\end{eqnarray}
with eigenvalues
\begin{equation}
E = \pm \sqrt{\left(
\cos(k) - \mu \right)^2 + \Delta^2 \sin^2(k) } \,.
\end{equation}
Rotating the basis, we write the Hamiltonian in the following off-diagonal form:
\begin{eqnarray}
\tilde H_1 (k)  =  \left(
\begin{array}{cc}
 0 & A (k) \\
 B (k) & 0 \\
 \end{array} \right) \,,\quad
 A(k) =  \left( \mu - \cos(k) \right)  + i\, \Delta \,\sin(k)  \,, \quad
B (k) =  \left( \mu - \cos(k)\right)  - i\, \Delta \,\sin(k)  \,.
\label{skew1S}
\end{eqnarray}

In the complex $k$-plane, the $EP$'s are given by
\begin{eqnarray}
k_{AK}^{\pm} = -i \ln \Big \lbrace
\frac{-\mu \pm \sqrt{ \Delta^2 + \mu^2-1}
}
{ \Delta -1 }
\Big \rbrace \,\mbox{ and }\,
k_{B K}^{\pm} = -i \ln \Big \lbrace
\frac{ \mu \pm \sqrt{ \Delta^2 + \mu^2 -1}
}
{ \Delta + 1}
\Big \rbrace \,,
\end{eqnarray}
corresponding to $A (k_{AK}^{\pm})=0$ and $ B(k_{B K}^{\pm}) =0 $ respectively. This is in conformity with our definition of $EP$'s given in the introduction. Here the subscript ``$K$'' to the wave vector index $k$ stands for Kitaev chain.

Expanding around the $EP$'s, we get:
\begin{eqnarray}
\tilde H_1 (k) &\simeq &
\left(
\begin{array}{cc}
 0 & 0 \\
 B (k_{AK}^{\pm}) & 0  \\
\end{array} \right)
+ \left(
\begin{array}{cc}
 0 & A'(k_{AK}^{\pm}) \\
 B'(k_{AK}^{\pm})  & 0  \\
\end{array} \right)
\, \left( k - k_{AK}^{\pm} \right)\,,\nn
\tilde H_{1} (k) &\simeq &
\left(
\begin{array}{cc}
 0 & A(k_{B K}^{\pm}) \\
 0 & 0  \\
\end{array} \right)
+ \left(
\begin{array}{cc}
 0 & A '(k_{B K}^{\pm}) \\
 B '(k_{B K} ^{\pm})  & 0  \\
\end{array} \right)
\, \left( k - k_{B K}^{\pm} \right)\,.\nn
\end{eqnarray}

The phase transitions occur at $ \mu = \pm 1$. For $|\mu| < 1$, we get a topological phase with one Majorana zero mode at each end of the chain. Moving from left to right along the $\mu$-axis, the transition are (1) from $0$ to $1$ chiral majorana mode as one crosses $\mu=-1 $, and (2) from $1$ to $0$ as one crosses $\mu = 1 $. We note the following:
\begin{enumerate}

\item At $\mu=- 1$, $\tilde H_1 (k_{AK}^+) = 0$ and $\tilde H_1 (k_{AK}^-) \neq 0$, while $\tilde H_1 (k_{B K} ^+) = 0$ and $ \tilde H_1 (k_{B K}^-) \neq 0$.

\item At $\mu= 1 $, $\tilde H_1 (k_{AK}^+) \neq 0$ and $ \tilde H_1 (k_{AK}^-) =0$, while $\tilde H_1 (k_{B K}^+) \neq 0$ and $ \tilde H_1 (k_{B K}^-) = 0$.

\item At all other values of $\mu$, $\tilde H_1 (k_{AK}^{\pm})$ and $\tilde H_1 (k_{B K}^{\pm})$ are non-zero.
\end{enumerate}
So, one of the two $EP$'s collapses at the phase transition points, irrespective of whether we consider $A(k)=0$ or $B(k)=0$.


\begin{figure}[ht]
\begin{center}
\subfigure[]{\includegraphics[scale=0.4]{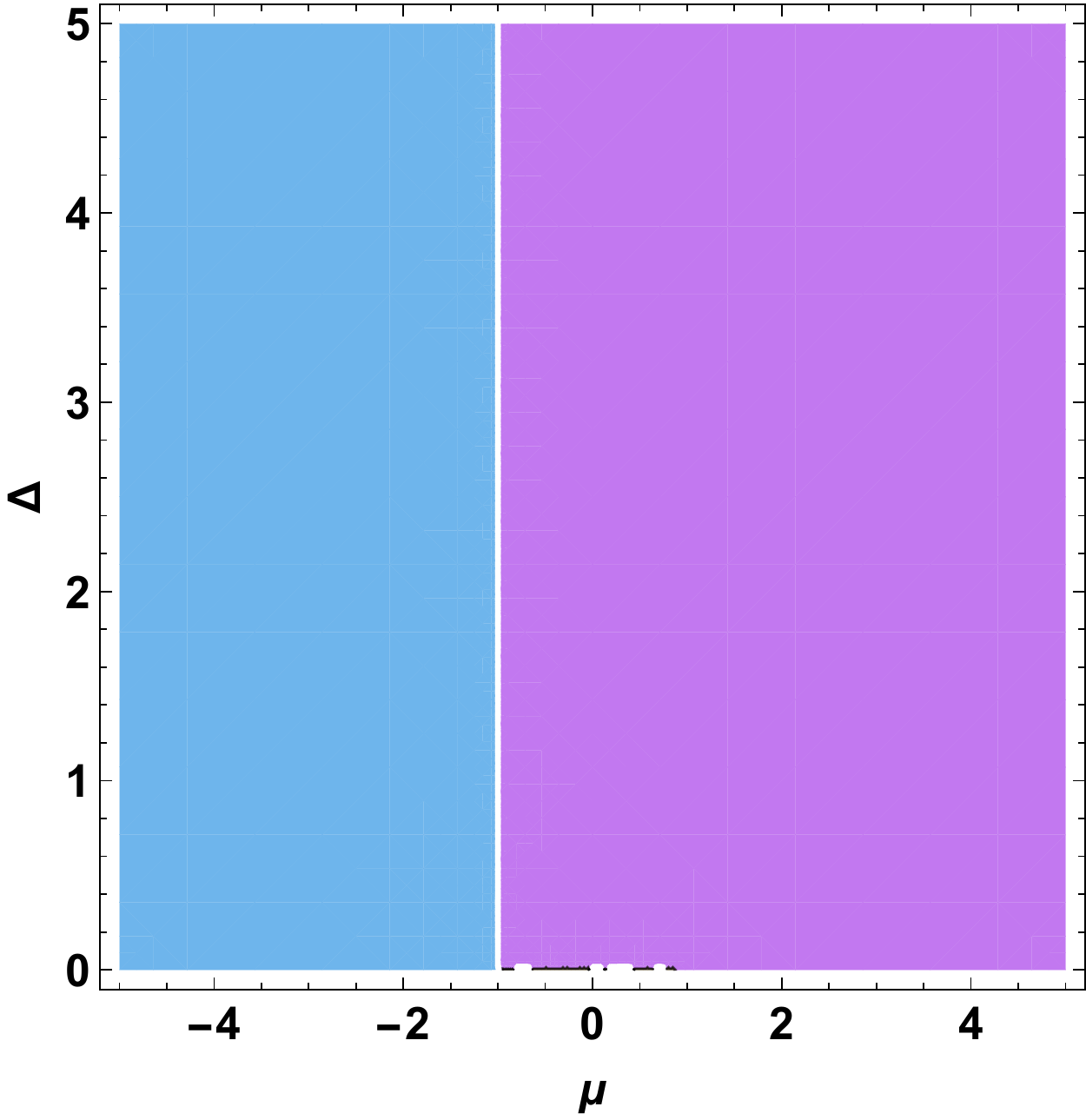}
\label{k1}}\qquad
\subfigure[]{\includegraphics[scale=0.4]{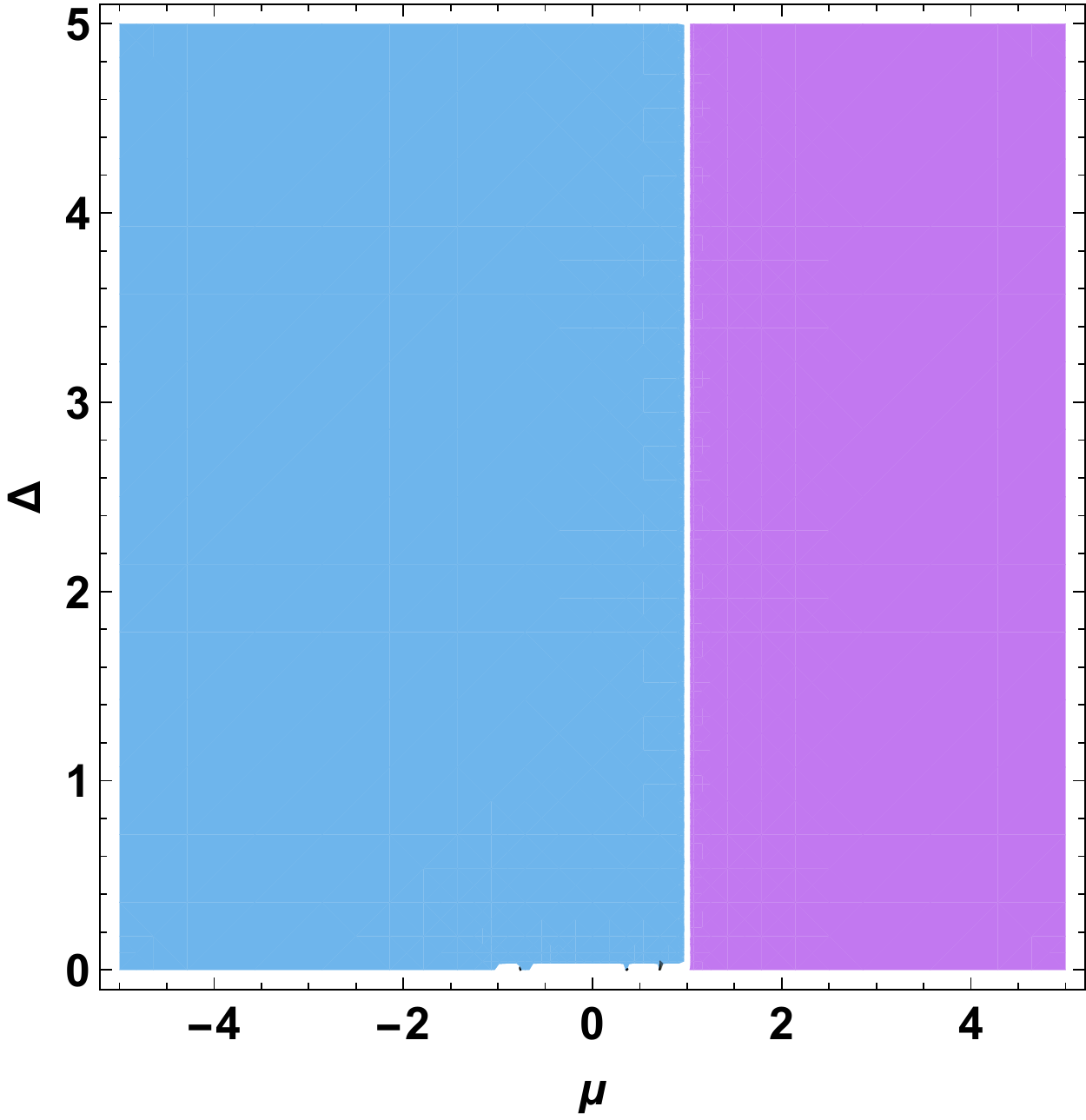}
\label{k2}}
\end{center}
\caption{(Color online)
(a) Blue region corresponds to $\Im \left ( k_{A K}^+ \right ) < 0 $ and $\Im \left ( k_{B K}^+ \right ) > 0 $. Purple region corresponds to $\Im \left ( k_{A K}^+ \right ) > 0 $ and $\Im \left ( k_{B K}^+ \right ) < 0 $.
(b) Blue region corresponds to $\Im \left ( k_{A K}^- \right ) > 0 $ and $\Im \left ( k_{B K}^- \right ) < 0 $. Purple region corresponds to $\Im \left ( k_{A K}^- \right ) < 0 $ and $\Im \left ( k_{B K}^- \right ) > 0 $.
}
\end{figure}

\begin{figure}[ht]
\centering
\includegraphics[scale=0.4]{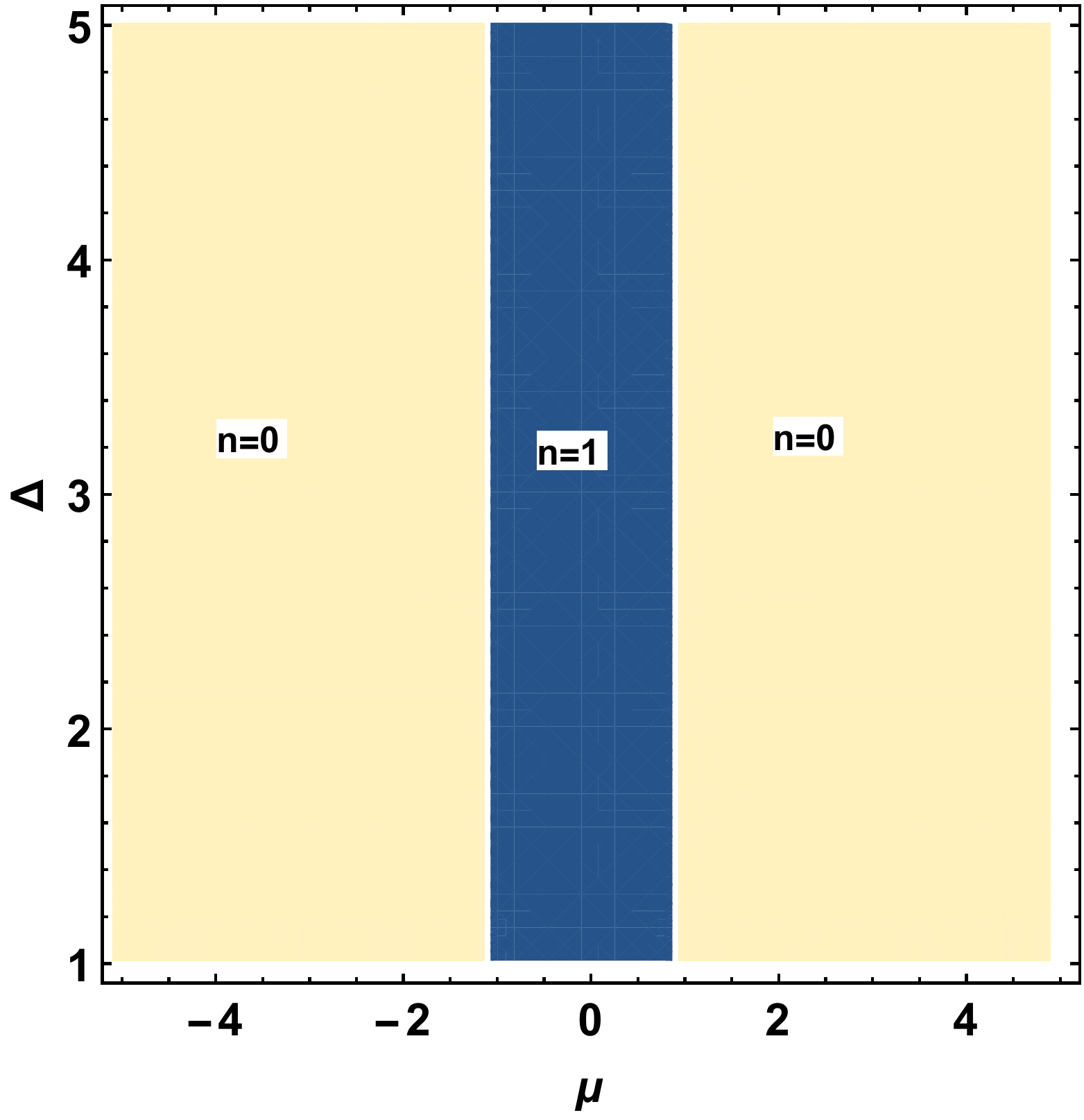}
\caption{\label{fig:f1}
(Color online) $f^{A,B} (\mu , \Delta)$ reproduces the topological phases of the Hamiltonian in Eq.~(\ref{h1}) in the $ ( \mu ,\Delta ) $-plane.
}
\end{figure}

Figs.~\ref{k1} and \ref{k2} show the dependence of the signs of $\Im \left ( k_{AK , BK}^{\pm} \right ) $ as functions of $(\mu , \Delta )$. At $\mu=-1$, we find that $ \Im \left ( k_{AK,BK}^+ (\mu ,\Delta ) \right ) $ change signs. For instance, $ \Im \left ( k_{AK}^+ (\mu ,\Delta ) \right ) < 0$ for $\mu <-1$, and $ \Im \left ( k_{AK}^+ (\mu ,\Delta ) \right ) > 0$ for $\mu >-1$. This means that $ \Im \left ( k_{AK}^+ (\mu ,\Delta ) \right ) = 0$ at $\mu =-1$, which in turn implies that $E(k)= 0$ has a solution for a real value of $k$ (as $E^2 (k) = A(k) \, B(k)$). Since $E(k)= 0$ for a real $k$ signals a topological quantum phase transition (gap-closing) in this system, we find that in the $EP$-description, the topological phase transition is marked by the imaginary component of one of the $EP$'s going through zero. We find similar behaviour at $\mu=1$. These observations help us define the functions:
\begin{eqnarray}
\label{fkitaev}
  f^{A,B} (\mu, \Delta)  
&=& \frac{1} {2} \,
\Big | sgn\big \lbrace \Im \left ( k_{AK,BK}^+ (\mu ,\Delta ) \right ) \big \rbrace
- sgn\big \lbrace \Im \left ( k_{AK,BK}^+(\mu_0 ,\Delta ) \right ) \big \rbrace \nn
&&
\, \quad +\,  sgn\big \lbrace \Im \left ( k_{AK,BK}^- (\mu,\Delta ) \right ) \big \rbrace
- sgn\big \lbrace \Im \left ( k_{AK , BK}^-(\mu_0 ,\Delta) \right ) \big \rbrace  \Big |
\,,\nn
\end{eqnarray}
which capture the number of chiral Majorana zero modes in a given phase. Here, $ \mu_0$ is any value of $\mu$ where we have a non-topological or zero Majorana mode phase, i.e.\ $\mu_0 \notin (-1,1) $.
Fig.~\ref{fig:f1} shows the contourplot for $f^{A,B} (\mu ,\Delta)$ in the $(\mu , \Delta)$-plane. In practice, one can work with either $f^A$ or $f^B $ to identify the topological phase.

These $EP$'s were studied in an earlier work \cite{sourin} in the context of topological phases for the Kitaev chain. The authors proposed that the function $\frac{1}{2} \Big[\,
 sgn\big \lbrace \Im \left ( k_{AK}^+ (\mu ,\Delta ) \right ) \big \rbrace
+ sgn\big \lbrace \Im \left ( k_{A K }^-(\mu ,\Delta ) \right ) \big \rbrace \Big ] $ gives the number of Majorana zero mode(s) in the various phases. However, we find that this definition works only for the $ 0 \rightarrow 1 $ Majorana fermion topological phase transition in the Kitaev model, and furthermore, when one chooses to work with the $EP$'s corresponding to $A (k)=0$ (and not for $B (k)=0$).


\section{Model $2$: Ising chain with longer-ranged interactions}
\label{model2}

The fermionized version of the $1D$ transverse field Ising model, generalized to include longer-ranged spin-spin interactions, can support $0$, $1$ or $2$ Majorana mode(s) at each end \cite{sudip}. The corresponding BdG Hamiltonian is given by
\begin{eqnarray}
\label{h2}
&& H_2 (k)  =  \left(
\begin{array}{cc}
 \xi(k) & i \, \Delta_I (k) \\
 -i \, \Delta_I (k) & -\xi(k) \\
 \end{array} \right) \,,\nn
&& \xi (k) = 2 \left(\,
 1- \lambda_1 \, \cos(k) -\lambda_2 \, \cos (2k)
 \, \right)\,, \quad
 \Delta_I (k) = 2 \left(\,
  \lambda_1 \, \sin(k) + \lambda_2 \, \sin (2k)
 \, \right)\,,
\end{eqnarray}
where $\lambda_1$ denotes the magnitudes of the nearest neighbour hopping and superconducting gap, and $\lambda_2$ denotes the amplitudes of the next nearest neighbour hopping and superconducting gap. It has been shown \cite{sudip} that the presence of $\lambda_2$ gives rise to two Majorana zero modes coexisting at the same end of the $1D$ chain, for certain values of the parameters $\lambda_1$ and $\lambda_2$. In this system, multiple zero energy Majorana modes can coexist (and do not mix and split to finite energies), because the Hamiltonian is in the chiral BDI class with an integer invariant \cite{TewariPRL2012}.The eigenvalues of the Hamiltonian in Eq.~(\ref{h2}) are given by
\begin{equation}
E_I(k) = \pm \sqrt{\xi^2(k) + \Delta_I^2(k) } \,.
\end{equation}
Since the Hamiltonian is chiral ($ \,\lbrace H_2(k), \sigma_x \rbrace  = 0$), by rotating the basis, we can rewrite it in the following off-diagonal form:
\begin{eqnarray}
\tilde H_{2} (k)  =  \left(
\begin{array}{cc}
 0 & A_I (k) \\
 B_I (k) & 0 \\
 \end{array} \right) \,,\quad
 A_I(k) &=&  \xi (k)  + i \,\Delta_I (k) \,, \quad
 B_I (k) =  \xi (k)  - i\, \Delta_I (k) \,.
\label{skew2}
\end{eqnarray}

In the complex $k$-plane, the $EP$'s are given by
\begin{eqnarray}
k_{AI}^{\pm} = -i \ln \Big \lbrace
\frac{-\lambda_1 \pm \sqrt{ \lambda_1^2 + 4\,\lambda_2}
}
{2 \, \lambda_2 }
\Big \rbrace \, \mbox{  and  } \,
k_{BI}^{\pm} = -i \ln \Big \lbrace
\frac{-\lambda_1 \pm \sqrt{ \lambda_1^2 + 4\,\lambda_2}
}
{2 }
\Big \rbrace \,,
\end{eqnarray}
corresponding to $A_I (k_{AI}^{\pm})=0$ and $ B_I(k_{BI}^{\pm}) =0 $ respectively.


The phase diagram for this model has been derived previously \cite{sudip}. Here we describe the topological phase diagram using the notion of the $EP$'s. Following the trajectories of the $EP$'s (derived by setting either $A_I (k)=0$ or $B_I (k)=0$) as functions of $\left(  \lambda_1, \lambda_2 \right)$, we find that along the phase transition lines from the $0$ to $2$ Majorana mode phase, $ \tilde H_{2} (k_{AI, BI}^{\pm}) = 0$, and hence both the $EP$'s collapse. However, along the transition lines from $0$ to $1$, or $1$ to $2$ Majorana zero modes, only one of the $EP$'s collapses, i.e.\ either $ \tilde H_{2} (k_{AI, BI}^{+})=0$ or $ \tilde H_{2} (k_{AI, BI}^{\pm}) =0$ (but not both).

\begin{figure}[ht]
\begin{center}
\subfigure[]{\includegraphics[scale=0.4]{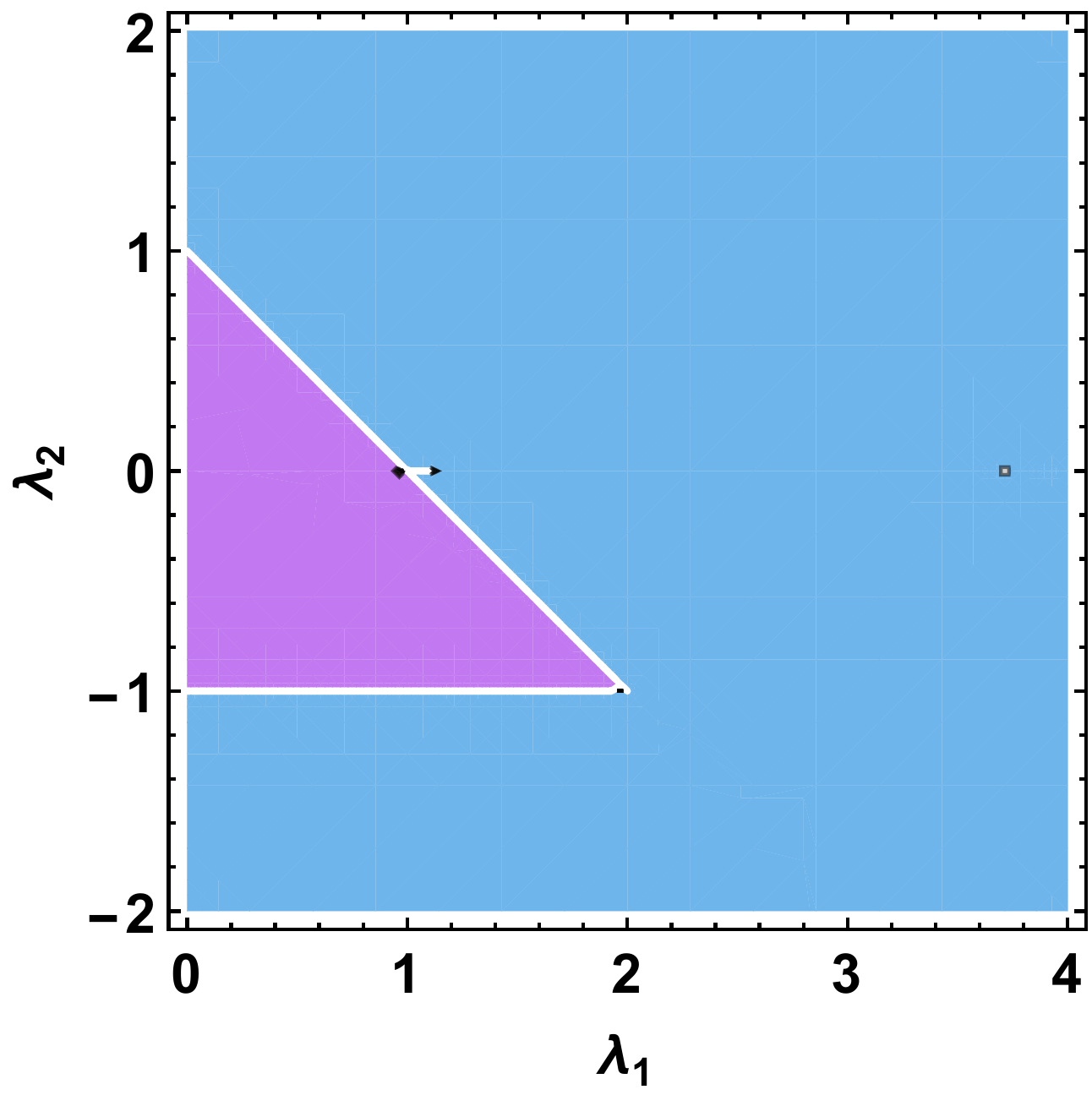}
\label{su1}}\qquad
\subfigure[]{\includegraphics[scale=0.4]{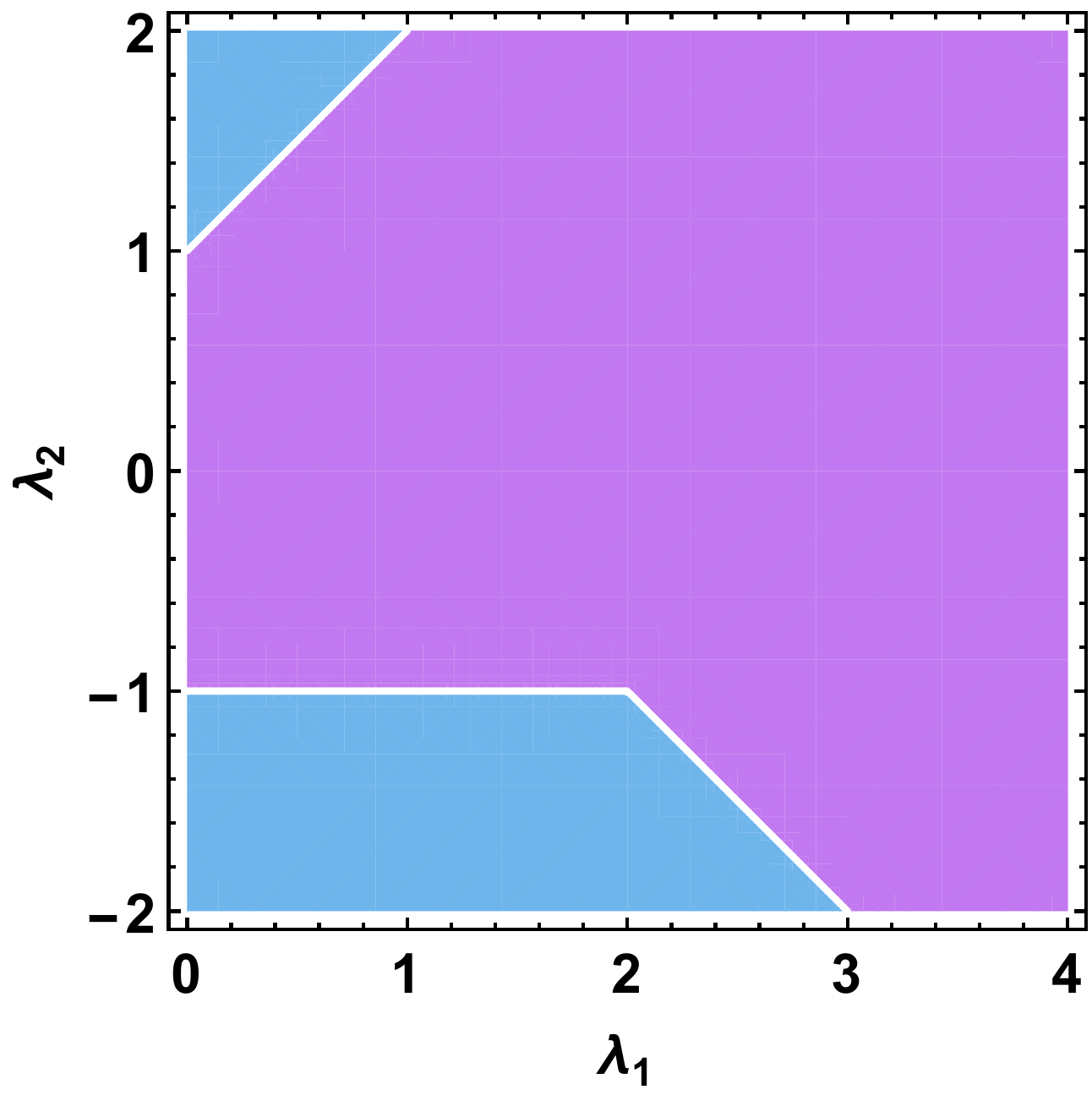}
\label{su2}}
\end{center}
\caption{(Color online)
(a) Blue region corresponds to $\Im \left ( k_{ AI }^+ \right ) > 0 $ and $\Im \left ( k_{BI }^+ \right ) < 0 $. Purple region corresponds to $\Im \left ( k_{ AI }^+ \right ) < 0 $ and $\Im \left ( k_{BI }^+ \right ) > 0 $.
(b) Blue region corresponds to $\Im \left ( k_{ AI }^- \right ) > 0 $ and $\Im \left ( k_{B I}^- \right ) < 0 $. Purple regions corresponds to $\Im \left ( k_{ AI }^- \right ) < 0 $ and $\Im \left ( k_{BI }^- \right ) > 0 $.
}
\end{figure}

\begin{figure}[ht]
\centering
\includegraphics[scale=0.3]{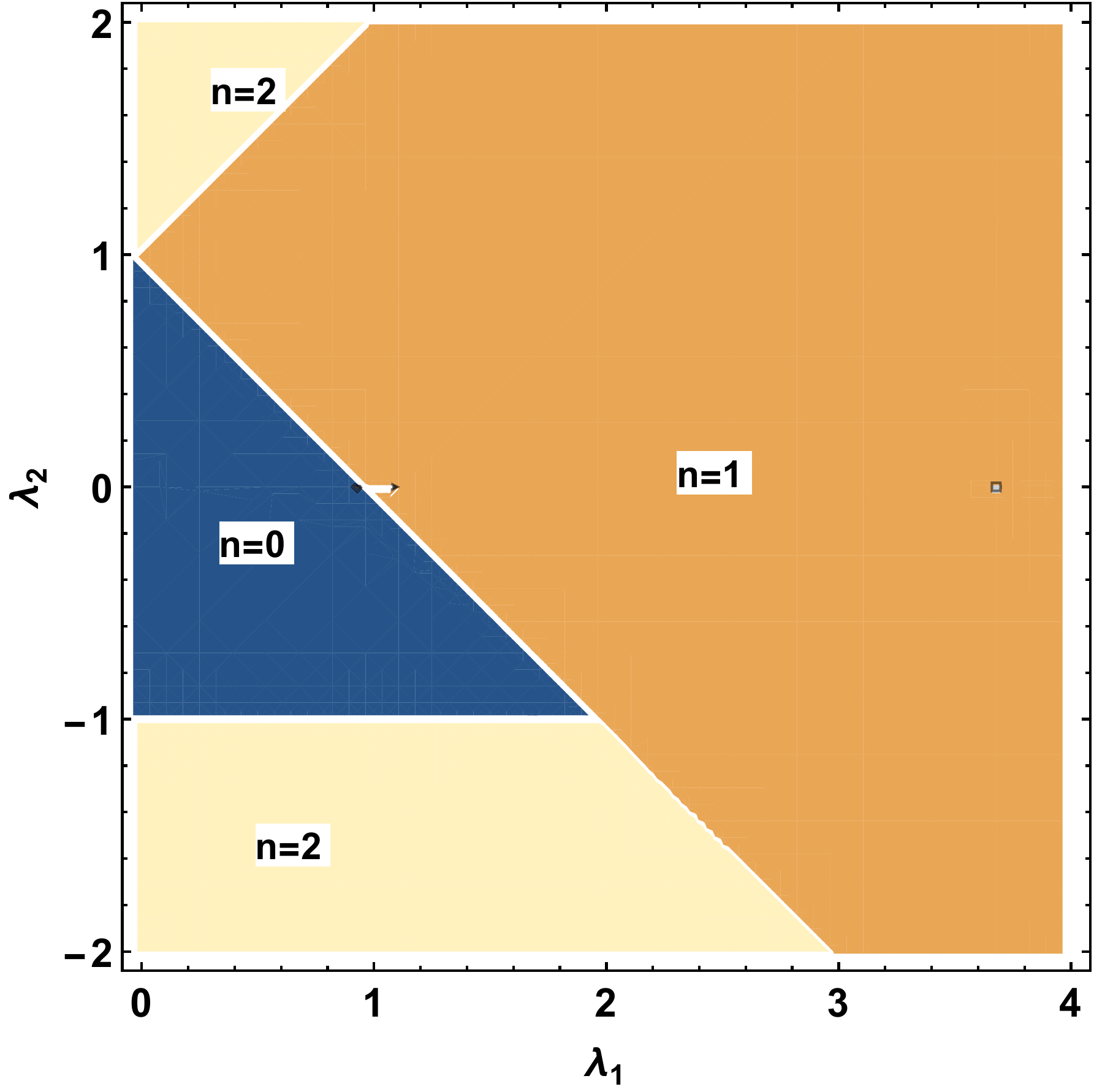}
\caption{(Color online) $f^{A,B} (\lambda_1, \lambda_2)$, defined in Eq.~(\ref{fising}), reproduces the topological phase diagram of the Hamiltonian in Eq.~(\ref{h2}), found earlier \cite{sudip} using the solutions of the BdG equations and the winding number topological invariant. Here, $n$ indicates the number of protected zero energy MBSs at any given end.}
\label{f2}
\end{figure}

Figs.~\ref{su1} and \ref{su2} show the dependence of the signs of $\Im \left ( k_{AI , BI }^{\pm} \right ) $ as functions of $(\lambda_1 , \lambda_2)$. Let $n$ denote the number of Majorana zero mode(s) at each end. Along the $ n=0 \rightarrow n=2 $ phase transition lines, both $\Im \left ( k_{AI,BI}^+ ( \lambda_1 , \lambda_2 ) \right )$ and $\Im \left ( k_{AI,BI}^- ( \lambda_1 , \lambda_2 ) \right )$ reduce to zero and go through a sign change. In Fig.~\ref{f2}, we show the contourplot for the functions
\begin{eqnarray}
\label{fising}
 f^{A,B} (\lambda_1 , \lambda_2 ) 
&=&  \frac{1} {2} \,
\Big | sgn\big \lbrace \Im \left ( k_{AI,BI}^+ ( \lambda_1 , \lambda_2 ) \right ) \big \rbrace
 - sgn\big \lbrace \Im \left ( k_{AI,BI }^+( \lambda_1^0 , \lambda^0_2 ) \right ) \big \rbrace \nn &&
\quad \,   + \, sgn\big \lbrace \Im \left ( k_{AI,BI}^- ( \lambda_1 , \lambda_2 ) \right ) \big \rbrace
- sgn\big \lbrace \Im \left ( k_{AI ,BI}^-(  \lambda_1^0 , \lambda_2^0 ) \right ) \big \rbrace 
\Big |\,,\nn
\end{eqnarray}
where $ (  \lambda_1^0 , \lambda_2^0 ) $ is any point in phase space where we have a non-topological phase with no Majorana zero mode. This reproduces the topological phase diagram for the Ising chain with longer-ranged interactions found earlier \cite{sudip}.


\section{Model $3$: Chiral Topological Ferromagnetic Nanowires}
\label{model3}

We consider the strictly $1D$ version of the Hamiltonian for a ferromagnetic nanowire embedded on Pb superconductor \cite{sumanta-chiral} with a single spatial channel (i.e.\ no transverse hopping). In the momentum space, the BdG Hamiltonian becomes $H=\sum_{k} \Psi_k^{\dagger}\, H_3 (k)\, \Psi_k$, where
\begin{eqnarray}
H_3 (k)  & = & \xi_n(k) \, \sigma_0  \,\tau_z 
+\left[ \Delta_s \, \sigma_0 + \Delta_p \sin(k)\, \vec{d} \cdot \vec{\sigma} \right] \tau_x 
	  + \, \vec{V} \cdot \vec{\sigma} \, \tau_0 \,,\quad
\xi_n(k) = -2t \cos(k) -\mu \,.
\label{h3}
\end{eqnarray}
Here $k \equiv k_x$ is the $1D$ crystal momentum, $\Psi_{k}=(c_{k\uparrow},c_{k\downarrow},c_{-k\downarrow}^{\dagger},-c_{-k\uparrow}^{\dagger})^{T}$ is the four-component Nambu spinor which acts on the particle-hole $(\vec \tau)$ and spin $(\vec \sigma)$ spaces, and $\vec{V}$ is the Zeeman field which can be induced by ferromagnetism. Also, $\Delta_s$ and $ \Delta_p $ are proximity-induced $s$-wave and $p$-wave superconducting pairing potentials respectively, with $\vec{d}$ determining the relative magnitudes of the components of the $p$-wave superconducting order parameter $\Delta_{\alpha \beta}$ $(\alpha,\beta = \uparrow , \downarrow )$.
In our calculations, we use $\vec{d}=(1,0,0)$ and $ \vec{V}=(0,V,0)$. Furthermore, we set $\Delta_s=0$, thus only considering $p$-wave pairing, which does not change the chiral BDI class of the Hamiltonian.

The eigenvalues of the Hamiltonian are given by:
\begin{eqnarray}
\label{e1e2}
E_1(k) = \pm \sqrt{ \left( V + \xi_n(k) \right)^2 + \Delta_p^2 \sin^2(k) } \,,\quad
E_2(k) = \pm \sqrt{ \left( V - \xi_n(k) \right)^2 + \Delta_p^2 \sin^2(k) } \,.
\end{eqnarray}
We change the basis to transform the Hamiltonian in Eq.~(\ref{h3}) to the form:
\begin{equation}
\tilde H_{3}(k) = \left(
\begin{array}{cc}
 0 & h_A(k) \\
 h_B(k) & 0 \\
 \end{array} \right)\,,\nn
\label{skew30}
\end{equation}
where
\begin{eqnarray}
&& h_A (k) = \left(
\begin{array}{cc}
 - \xi_n(k) + i \, \Delta_p \sin(k) & V \\
 - V & \xi_n(k)  - i \, \Delta_p \sin(k) \\
\end{array} \right)\,, \quad
h_B (k) = \left(
\begin{array}{cc}
 - \xi_n(k) - i \, \Delta_p \sin(k) & - V \\
 V & \xi_n(k)  + i \, \Delta_p \sin(k) \\
\end{array} \right)\,.\nn
\label{skew3}
\end{eqnarray}
This corresponds to the chiral basis where the chiral operator $ S = \sigma_x \, \tau_y $ is rotated to the diagonal
form $ \text{diag} (-1,-1,1,1 ) $.
The upper block $h_A(k )$ and the lower block $h_B(k)$ of $H_{3}(k) $ give eigenfunctions of opposite chirality with respect to $ S $.

In the complex $k$-plane,
the $EP$'s for chirality solutions are given by
\begin{eqnarray}
k_{A,1}^{\pm} = -i \ln \Big \lbrace
\frac{ V -\mu \pm \sqrt{ \left( V - \mu \right)^2 + \Delta_p^2 - 4 t^2}
}
{2 t + \Delta_p}
\Big \rbrace \,, \quad
k_{A,2}^{\pm} = -i \ln \Big \lbrace
\frac{-V -\mu \pm \sqrt{ \left( V + \mu \right)^2 + \Delta_p^2 - 4 t^2}
}
{2 t + \Delta_p}
\Big \rbrace \,, 
\end{eqnarray}
corresponding to $  \det \big[ h_A (k) \big] =0 $.
Similarly, the $EP$'s for negative chirality solutions are given by
\begin{eqnarray}
k_{B,1}^{\pm} = -i \ln \Big \lbrace
\frac{ V -\mu \pm \sqrt{ \left( V - \mu \right)^2 + \Delta_p^2 - 4 t^2}
}
{2 t - \Delta_p}
\Big \rbrace \,, \quad
k_{B,2}^{\pm} = -i \ln \Big \lbrace
\frac{-V -\mu \pm \sqrt{ \left( V + \mu \right)^2 + \Delta_p^2 - 4 t^2}
}
{2 t - \Delta_p}
\Big \rbrace \,, 
\end{eqnarray}
corresponding to $  \det \big[ h_B (k) \big] =0 $. In either case, $H_3(k) $ becomes non-diagonalizable.


\begin{figure}[ht]
\begin{center}
\includegraphics[scale=0.4]{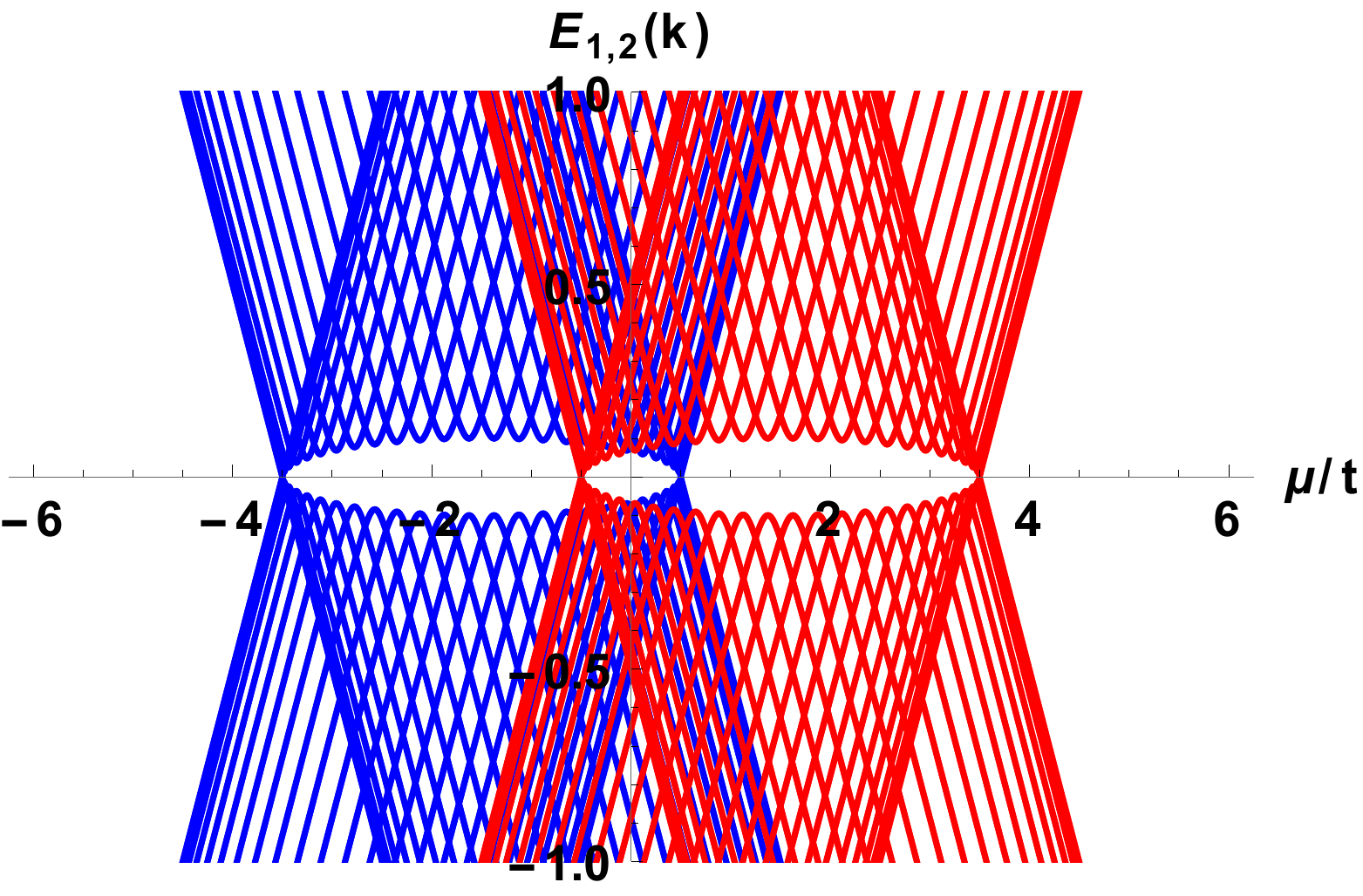}
\end{center}
\caption{(Color online)
Energy bands $E_{1,2} (k)$, given in Eq.~(\ref{e1e2}), have been plotted in red and blue respectively as functions of $\mu/t$. We have used the values $V= 1.5 \,t$ and $ \Delta_p=0.1 \, t$.
}
\label{chiral-spec}
\end{figure}

Choosing $ V= 1.5 \,t\,, \Delta_p = 0.1 \, t $, the energy bands are shown in Fig.~\ref{chiral-spec} as functions of $\mu/t$. So, there are phase transitions at $\mu= \lbrace -3.5\,t , 0.5 \, t\rbrace$ for $h_2(k)$, and at $\mu= \lbrace -0.5\,t , 3.5 \, t\rbrace$ for $h_1(k)$. Moving from left to right along the $\mu$-axis, the transitions are (1) from $0$ to $1$ chiral Majorana mode as one crosses $\mu=-3.5 \, t$, and (2) from $1$ to $2$ chiral Majorana modes as one crosses $\mu = - 0.5 \, t$. Let us examine the $EP$'s corresponding to $\det \big[ h_A (k) \big] =0$:
\begin{enumerate}

\item At $\mu= - 0.5 \, t$, $ \det \big[ h_B (k_{A,1} ^+) \big] =0 $, while $\det \big[ h_B (k_{A,1}^-) \big] $ and
$ \det \big[ h_B (k_{A,2}^\pm) \big ] $ are non-zero. Hence, $H_3 (k_{A,1} ^+)$ is diagonalizable indicating the collapse of the $EP$ at $k= k_{A,1} ^+ $.

\item At $\mu= 3.5 \, t$, $ \det \big[ h_B (k_{A,1} ^-) \big] =0 $, while $\det \big[ h_B (k_{A,1}^+) \big] $ and
$ \det \big[ h_B (k_{A,2}^\pm) \big ] $ are non-zero. $H_3 (k_{A,1} ^-)$ is thus diagonalizable indicating the collapse of the $EP$ at $k= k_{A,1} ^- $.

\item At $\mu=- 3.5 \, t$, $ \det \big[ h_B (k_{A,2} ^+) \big] =0 $, while $\det \big[ h_B (k_{A,2}^-) \big] $ and
$ \det \big[ h_B (k_{A,1}^\pm) \big ] $ are non-zero. $H_3 (k_{A,2} ^+)$ is thus diagonalizable indicating the collapse of the $EP$ at $k= k_{A,2} ^+ $.

\item At $\mu= 0.5 \, t$, $ \det \big[ h_B (k_{A,2} ^-) \big] =0 $, while $\det \big[ h_B (k_{A,2}^+) \big] $ and
$ \det \big[ h_B (k_{A,1}^\pm) \big ] $ are non-zero. $H_3 (k_{A,2} ^-)$ is thus diagonalizable indicating the collapse of the $EP$ at $k= k_{A,2} ^- $.

\end{enumerate}
Similar observations hold if we consider the $EP$'s corresponding to $\det \big[ h_B (k) \big] =0$.
So, one of the four $EP$'s collapses at each phase transition point, irrespective of whether we decide to examine the $\det \big[ h_A (k) \big] =0$ or $ \det \big[ h_B (k) \big] =0$ solutions.

\begin{figure}[ht]
\begin{center}
\subfigure[]{\includegraphics[scale=0.4]{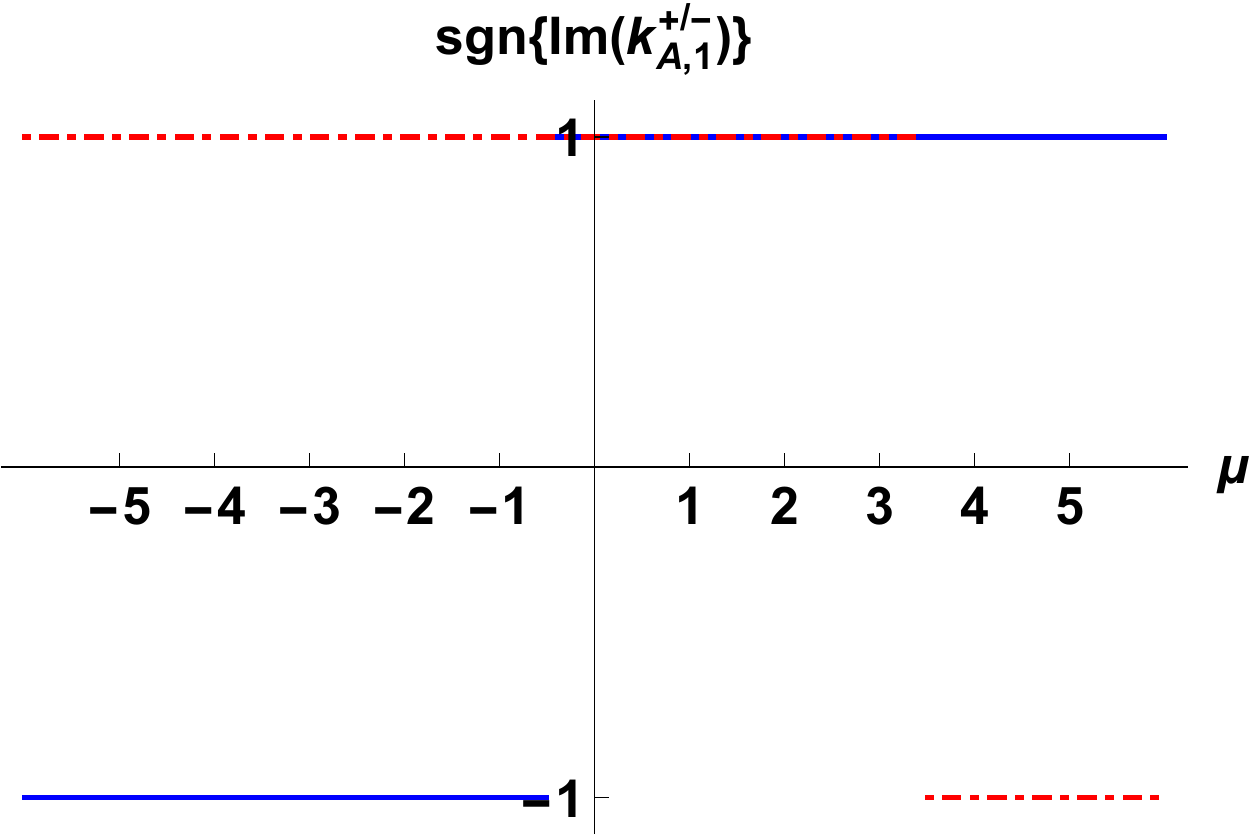}
\label{ka1}}
\subfigure[]{\includegraphics[scale=0.4]{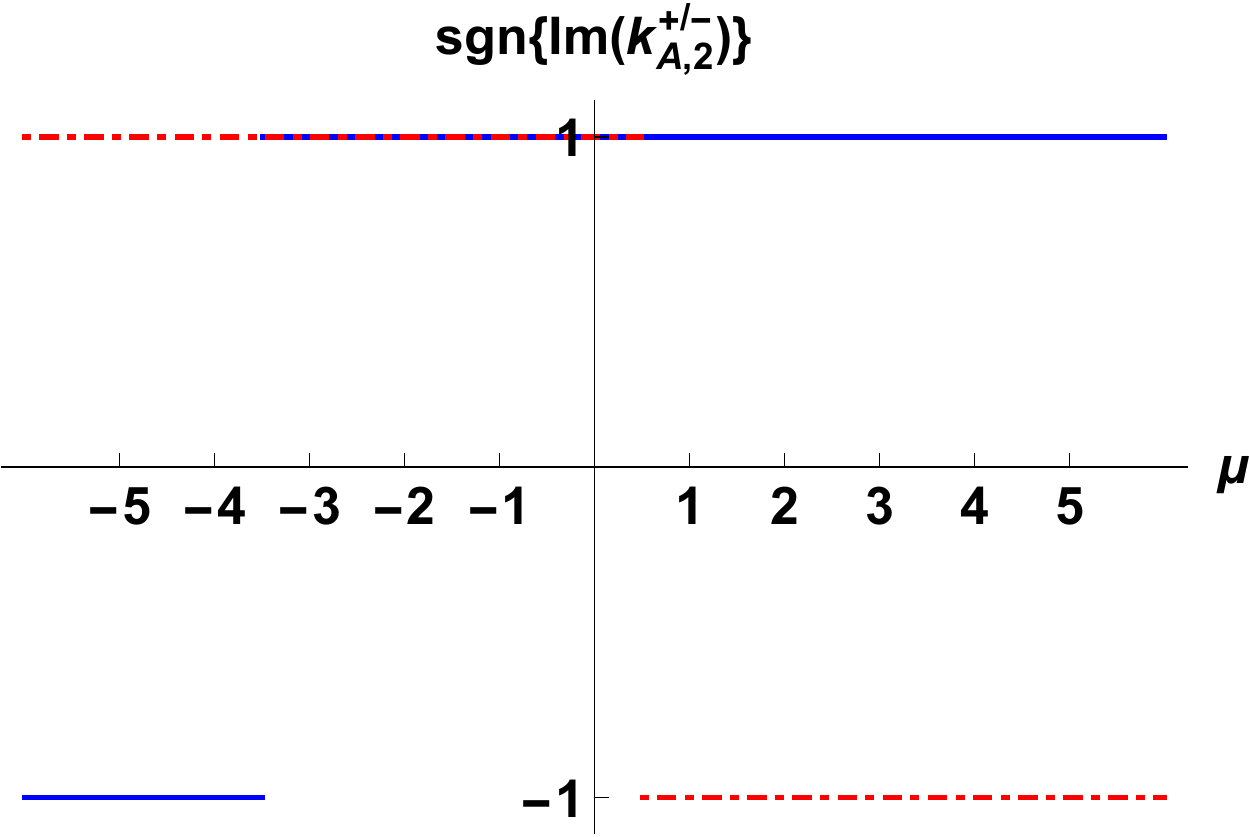}
\label{ka2}} \\
\subfigure[]{\includegraphics[scale=0.4]{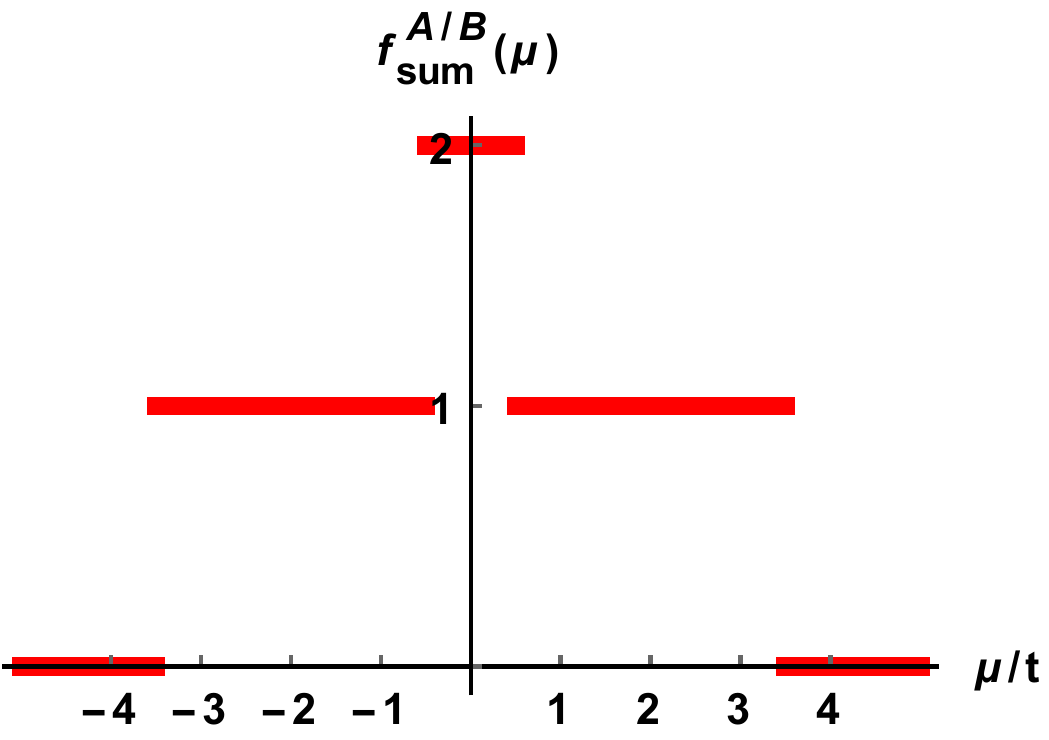}
\label{f3}} \\
\end{center}
\caption{(Color online)
Parameters: $V= 1.5 \,t, \, \Delta_p= 0.1 \, t $.
(a) Blue and dotted red lines correspond to  $ sgn\big \lbrace \Im \left ( k_{A,1}^+  \right ) \big \rbrace $ and $ sgn\big \lbrace \Im \left ( k_{A,1}^- \right ) \big \rbrace $ respectively.
(b) Blue and dotted red lines correspond to  $ sgn\big \lbrace \Im \left ( k_{ A,2 }^+  \right ) \big \rbrace $ and $ sgn\big \lbrace \Im \left ( k_{ A,2 }^- \right ) \big \rbrace $ respectively.
(c) $f_{sum}^{A/B}(\mu)$ giving the count of the chiral Majorana zero modes as functions of $\mu/t$.
}
\end{figure}

Figs.~\ref{ka1} and \ref{ka2} show the dependence of the signs of $\Im \left ( k_{A, 1}^{\pm} \right ) $ and $\Im \left ( k_{A ,2}^{\pm} \right ) $ on $\mu /t $. Generalizing the functions defined in Eqs.~(\ref{fkitaev}) and (\ref{fising}) to the present case of four $EP$'s, we define:
\begin{eqnarray}
 f_{sum}^{A/B} (\mu) &=&  \frac{1} {2} \,\Big | \sum_{s= \pm } \sum_{r=1,2 } 
\Big[\,
 sgn\big \lbrace \Im \left ( k_{A/B,r}^s ( \mu  ) \right ) \big \rbrace
 - sgn\big \lbrace \Im \left ( k_{ A/B,r }^s( \mu_0  ) \right ) \big \rbrace 
\Big ] \Big | \,,
\end{eqnarray}
which captures the number of chiral Majorana zero modes in a given phase. Here, $\mu_0$ is any value of $\mu$ where we have a non-topological phase. Fig.~\ref{f3} shows the plot for $f_{sum}^{A/B} (\mu)$.

\section{Generic formula}
\label{gen-formula}


A generic $1D$ $ 2N \times 2N $ chiral Hamiltonian $ H_{\textsl{chiral}} (k) $ can support multiple Majorana zero modes at each end of the fermionic chain. Let $k=k_{EP}^{j}$ ( for $ j = 1, 2 , \ldots, n_{ep}$)  be the solutions of the $EP$'s corresponding to the vanishing of the determinant of any one off-diagonal block, after $ H_{\textsl{chiral}}$ has been rotated into the off-diagonal form and $k$ has been promoted to a complex number.
From the study of our models, we observe the following:
\begin{enumerate}
\item  In Sec.~\ref{model1}, $ sgn \big \lbrace \Im \left ( k_{A K}^+ ( \lbrace p_i \rbrace  ) \right ) \big \rbrace = sgn \big \lbrace \Im \left ( k_{ A K}^- ( \lbrace p_i \rbrace  ) \right ) \big \rbrace $ for a phase with $n=1$ zero mode. Furthermore, $ sgn \big \lbrace \Im \left ( k_{ A K }^+ ( \lbrace p_i \rbrace  ) \right ) \big \rbrace $ and $sgn \big \lbrace \Im \left ( k_{ A K }^- ( \lbrace p_i \rbrace  ) \right ) \big \rbrace $ have opposite signs for the non-topological phases.

\item   In Sec.~\ref{model2}, $ sgn \big \lbrace \Im \left ( k_{A I}^+ ( \lbrace p_i \rbrace  ) \right ) \big \rbrace = sgn \big \lbrace \Im \left ( k_{ A I }^- ( \lbrace p_i \rbrace  ) \right ) \big \rbrace $ for a phase with $n=0,2$ zero modes, and these signs are opposite to each other for the $n=0$ and $n=2$ cases. Furthermore, $ sgn \big \lbrace \Im \left ( k_{A I}^+ ( \lbrace p_i \rbrace  ) \right ) \big \rbrace $ and $sgn \big \lbrace \Im \left ( k_{A I }^- ( \lbrace p_i \rbrace  ) \right ) \big \rbrace $ have opposite signs for a phase with $n= 1$ zero mode.

\item In Sec.~\ref{model3}, one of the four $EP$'s undergo sign change at each phase transition point, and the sign dependence can be examined from Figs.~\ref{ka1} and \ref{ka2}.

\end{enumerate}
Here we discussed only the results for the upper off-diagonal block. But similar results hold for the lower one.

Hence, we propose the following generic formula:
\begin{eqnarray}
\label{fchiral}
f  (\mu) &=&  \frac{1} {2} \,\Big |  \sum_{j=1}^ { n_{ep } } 
\Big[\,
 sgn\big \lbrace \Im \left ( k_{EP}^j ( \lbrace p_i \rbrace  ) \right ) \big \rbrace 
 - sgn\big \lbrace \Im \left ( k_{ EP }^j( \lbrace p_i^0 \rbrace  ) \right ) \big \rbrace 
\Big ] \Big | \,,
\end{eqnarray}
where $k_{EP}^{j}$ (for $j=1,2 , \ldots n_{ep}$) are the $EP$ solutions found by solving any one of the off-diagonal blocks, $ \lbrace p_i \rbrace $ is the set of parameters appearing in the expressions for $k_{EP}^{j}$, and $ \lbrace p_i^0 \rbrace $ are their values at any point in the non-topological phase.

We emphasize that we can apply this formula for any chiral system in $1D$, i.e.\ for any chiral topological superconductor in the BDI class with an arbitrary integer topological invariant $\mathbb{Z}$. One example is the the case of several nanowires coupled by a transverse hopping term.
For the BDI class in $1D$, MBSs appearing at one end are of the same chirality, which at the same time characterize the winding number in that phase. Hence it is sufficient to consider the $EP$'s corresponding to the solutions of one of the off-diagonal blocks in order to apply Eq.~(\ref{fchiral}).

We now prove why one (or more) $\Im \left ( k_{EP}^{j} \right ) $ changes sign at a topological phase transition point, characterised by a set of parameters $\lbrace p_i = p_i^t\rbrace $. We consider a $1D$ $ 2 N \times 2  N$ Hamiltonian
\begin{eqnarray}
&& H_{od} (k)  =  \left(
\begin{array}{cc}
0 & h_A(k,\lbrace p_i\rbrace) \\
h_B(k,\lbrace p_i\rbrace)  & 0 \\
 \end{array} \right) \,,
\label{skew}
\end{eqnarray}
after rotating it to the off-diagonal form, such that the $N \times N$ matrices $h_A$ and $h_B$ satisfy $h_B (k,\lbrace p_i\rbrace) =h_A^{\dagger} (k,\lbrace p_i\rbrace)$ for real (physical) values of $k$.
Let
\begin{equation}
\det \big [ h_A (k,\lbrace p_i\rbrace) \big ] = a(k ,\lbrace p_i\rbrace) - i \,b (k, \lbrace p_i\rbrace) \,,
\end{equation}
where the parameters $\lbrace p_i\rbrace$ are real, and the functions $(a, \,b)$ are real for real $k$.
Then it follows that
\begin{equation}
\det \big [ h_B (k,\lbrace p_i\rbrace) \big ] = a(k ,\lbrace p_i\rbrace) + i \,b (k, \lbrace p_i\rbrace) \,.
\end{equation}
Let us consider the $EP$'s at $k=k_{EP}$ in the complex $k$-plane corresponding to $\det \big [ h_A (k,\lbrace p_i\rbrace) \big ] =0  \Rightarrow a(k_{EP},\lbrace p_i\rbrace) =i \,b (k_{EP}, \lbrace p_i\rbrace) $.
Let
\begin{equation}
k_{EP} = x (\lbrace p_i\rbrace )+ iy (\lbrace p_i\rbrace)\,,
\end{equation}
where $(x,y)$ are the real and imaginary parts of $k_{EP}$ as functions of $\lbrace p_i\rbrace$.
Then we must have
\begin{eqnarray}
a(k_{EP}, \lbrace p_i\rbrace) = f_a^{even} (x,y) + i f_a^{odd} (x,y) \,,\quad
b(k_{EP}, \lbrace p_i\rbrace) = f_b^{even} (x,y) + i f_b^{odd} (x,y) \,,
\end{eqnarray}
such that $f^{even}_{a,b}(x,y)$ and $f^{odd}_{a,b}(x,y)$ are even and odd functions of $y$ respectively. Needless to add that they must be real functions too. Now the constraint $ a(k_{EP},\lbrace p_i\rbrace) =i \,b (k_{EP}, \lbrace p_i\rbrace) $
translates into the equations
\begin{eqnarray}
f_a^{even} (x,y) &=& - f_b^{odd} (x,y) \,,\quad
f_a^{odd} (x,y) = f_b^{even} (x,y) \, ,
\end{eqnarray}
whose solutions give $(x,y)$ as functions of $\lbrace p_i\rbrace $. At a point $ \lbrace p_i = p_i^t\rbrace$ in parameter space, where both $a \pm ib$ vanish (i.e.\ $\det \big [ h_A\big ] = \det \big [ h_B \big ] = 0 $), we must have
\begin{eqnarray}
a(\lbrace p_i^t\rbrace) = b(\lbrace p_i^t\rbrace)=0   \quad
\Rightarrow f_a^{odd} (x,y)|_{\lbrace p_i=p_i^t \rbrace} = f_b^{odd} (x,y)|_{\lbrace p_i=p_i^t\rbrace} =0\, \quad
\Rightarrow y(\lbrace p_i^t\rbrace) = 0
\end{eqnarray}
as one of the solutions for $y(\lbrace p_i^t\rbrace)$.
This is clearly a topological phase transition point, as $ \det \big [ H_{od} (k) \big ] $ vanishes for a real value of $k$, which indicates one or more energy eigenvalues going to zero. At this point, the $EP$ also collapses as the Hamiltonian becomes diagonalizable.
For a point $ \lbrace p_1, p_{i \neq 1}^t \rbrace $ close to $ \lbrace p_1^t, p_{i \neq 1}^t \rbrace $, we have the expansion
\begin{equation}
y (\lbrace p_1, p_{i \neq 1}^t \rbrace ) =
\begin{cases} |p_1 - p_1^t| \, \partial_{p_1^t} y(\lbrace p_i^t\rbrace) &\mbox{if } p_1 > p_1^t \,,\\
- |p_1 - p_1^t| \, \partial_{p_1^t} y(\lbrace p_i^t\rbrace) &\mbox{if } p_1 < p_1^t \,, \end{cases}
\end{equation}
for the solution $y(\lbrace p_i^t\rbrace) = 0$.
Hence, this solution for $\Im \left ( k_{EP} \right ) $ undergoes a sign change on crossing the phase transition point.

Our counting formula is supplementary to other counting methods studied in the literature, like winding number calculation \cite{Schnyder_2008,Ryu_2010,TewariPRL2012}, scattering matrix approach \cite{scattering1,scattering2} and gradient expansion \cite{luiz}. The advantage of using our formula is that one has to solve only for the zeroes of the determinant of one off-diagonal block in terms of complex $k$. One need not perform any integral or other lengthy computations to find the number of MBSs characterizing the topological phase. Hence, in spite of conveying the same physics, we believe this method is slightly easier and more convenient to implement than the other standard methods.

From our entire analysis, we can intuitively understand that these $EP$ solutions in the complex $k$-plane can be mapped to the MBS wavefunctions in the real space with open boundary conditions. At any one end of a chiral $1D$ system, only MBS wavefunctions of a definite chirality can appear, which is reflected by the fact that we need to solve for $EP$'s only for one off-diagonal block and this gives us the number of MBSs at each end of an open chain. Again, the fact that $\Im \left ( k_{EP} \right )$ changes sign at a topological phase transition is tied to the fact that an MBS mode appears or disappears on moving from one phase to the other. One can prove the correspondence $\exp \left (i \, k_{EP}\, x \right)\leftrightarrow \exp \left( - z \, x \right )$, where $\exp \left( - z \, x \right )$ represents an exponentially decaying MBS wavefunction at a distance $x$ from the edge. Thus the imaginary part of momentum at the exceptional point indeed gives us the exponential decay of the corresponding MBS in real space. A rigorous proof of this bulk-boundary correspondence has been provided in subsequent works \cite{ipsita1,ipsita2}.

We note that we cannot apply this $EP$ formalism to count the Majorana zero modes for $1D$ Hamiltonians in class D. This is because a chiral symmetry operator does not exist and the Hamiltonian cannot be unitarily rotated into the block-diagonal form. It can be represented as:
\begin{eqnarray}
&& H_{D} (k) 
=  
\cos \theta 
\left(
\begin{array}{cc}
 0 & h_A(k,\lbrace p_i\rbrace) \\
h_B(k,\lbrace p_i\rbrace)  & 0 \\
 \end{array} \right)
  +
 \sin \theta
\left(
\begin{array}{cc}
\mathbb{I} & 0  \\
 0  &   - \mathbb{I}   \\
 \end{array} \right)
  \,,
\label{classd}
\end{eqnarray} 
with a non-vanishing value of $\sin \theta$.
In such a situation, the $EP$ solutions for the Hamiltonian will not correspond to the vanishing energy eigenvalues in the complex $k$-plane. For class DIII systems, the Majorana fermions at one end of the edge are not of the same chirality, but correspond to Majorana
Kramers pairs (MKPs) which are doubly degenerate Majorana zero modes. So to count those modes, we have to apply some additional criterion to the counting formula. The counting formula for the cases of class D and DIII systems has been discussed  in subsequent works \cite{ipsita1,ipsita2}.

\section{Conclusion}
\label{conclusion}

We have derived a generic formula for counting the number of Majorana zero modes for a $1D$ chiral topological superconductor/superfluid. First we write the Hamiltonian in the basis where the chiral symmetry opertaor is diagonal. In this basis, the Hamiltonian consists of two off-diagonal blocks. The solutions, $ k_{EP}^{j}$ (for $j=1,2 , \ldots n_{ep}$), for the exceptional points in the complex momentum space, can be obtained in terms of the parameters of the Hamiltonian by setting the determinant of any one off-diagonal block to zero. Our formula is based on the evolution of these $EP$'s in the complex $k$-plane as functions of the parameters. The count of the MBSs is encoded in the signs of $\Im \left ( k_{EP}^{j} \right ) $. At an $EP$, by definition, the complexified Hamiltonian is non-diagonalizable, due to one of the eigenvectors having a vanishing norm. However, at a physical phase transition point, at least one of the $ k_{EP}^j $'s becomes real making the Hamiltonian diagonalizable. This is due to the vanishing of the determinants of both the off-diagonal blocks, thus signalling the collapse of the corresponding $EP$.


\section{Acknowledgments}
We thank Atri Bhattacharya, Jay D. Sau and Sourin Das for stimulating discussions. We are also grateful to Chen-Hsuan Hsu for his valuable comments on the manuscript. I.M. was partially supported by the Templeton Foundation.
Research at the Perimeter Institute is supported
in part by the Government of Canada
through Industry Canada,
and by the Province of Ontario through the
Ministry of Research and Information. Research at Clemson is supported by 
the grant AFOSR (FA9550-13-1-0045).

\bibliography{exceptional-paper}

\appendix

\section{Review of \textit{Exceptional Points}}
\label{app:append}
``Exceptional Points" are branch point singularities in the parameter space of a matrix, at which at least two or more eigenvalues coincide. These are different from the familiar degeneracy observed for a Hermitian operator because of the fact that both the eigenvalues and the corresponding eigenstates coalesce \cite{Heiss}. Since a Hermitian operator has a complete set of eigenstates, $EP$'s can occur only in the spectrum of a non-Hermitian operator. Hence, if we analytically continue a real parameter of a Hamiltonian to complex values, the resulting system becomes non-Hermitian and we may find values of that complexified parameter where $EP$'s appear. Let us demonstrate the appearance of the $EP$'s for the simplest case of a $2 \times 2$ matrix:
\beq
\mathcal{H}  (\gamma) =
\left(
\begin{array}{cc}
 \mathcal{E}_1 & 0 \\
 0 & \mathcal{E}_2 \\
\end{array} \right)
+ \gamma 
\left(
\begin{array}{cc}
 \epsilon_1 &  \eta_1 \\
 \eta_2  & \epsilon_2  \\
\end{array} \right)  .
\eeq
The eigenvalues of $\mathcal{H} $ are given by:
\beq
E =
\frac{1}{2}
\Big \lbrace
\mathcal{E}_1 + \mathcal{E}_2
+ \gamma (\epsilon_1 + \epsilon_2)
\pm \sqrt{  (\mathcal{E}_1 - \mathcal{E}_2 + \epsilon_1 \gamma - \epsilon_2 \gamma )^2 
+ 4\, \gamma^2 \, \eta_1  \eta_2  } \,
\Big \rbrace \,,
\eeq
which coalesce at
\beq
(\mathcal{E}_1 - \mathcal{E}_2 + \epsilon_1 \gamma - \epsilon_2 \gamma )^2 
+ 4\, \gamma^2 \, \eta_1  \eta_2 =0
\Rightarrow
\gamma \equiv \gamma_\pm
= -
\frac{i \, ( \mathcal{E}_1 - \mathcal{E}_2 ) }
{i \, (\epsilon_1 - \epsilon_2) \pm \sqrt{\eta_1  \eta_2 }  } \,.
\eeq
At these points, $\mathcal{H}$ is a non-diagonalizable non-Hermitian matrix, which is manifested by the fact that there is only one linearly independent eigenvector (instead of two) and with a vanishing norm. The bi-orthogonal system for a non-Hermitian operator gives the right and left eigenstates proportional to:
\bqa
&& \left(
\begin{array}{cc}
 \frac{i \, \eta_1} { \sqrt{  \eta_1 \, \eta_2}  }  \\
 1    \\
\end{array} \right) \,,
\quad ( \frac {i \, \eta_2 } { \sqrt{ \eta_1 \, \eta_2} } , 1)  
\quad \mbox{ for } \gamma = \gamma_+ \,;\nn
&& \left(
\begin{array}{cc}
 \frac{  - i \, \eta_1} { \sqrt{  \eta_1 \, \eta_2}  }  \\
 1    \\
\end{array} \right) \,,
\quad ( \frac {  - i \, \eta_2 } { \sqrt{ \eta_1 \, \eta_2} } , 1) 
\quad  \mbox{ for } \gamma = \gamma_- \,,
\eqa
which clearly have zero norm.
Generalization to a higher dimensional matrix is straightforward.


\end{document}